\documentclass[acmtog]{acmart}

\usepackage{multirow}

\usepackage[ruled]{algorithm2e}

\SetAlFnt{\small}
\SetAlCapFnt{\small}
\SetAlCapNameFnt{\small}
\SetAlCapHSkip{0pt}

\acmSubmissionID{222}

\copyrightyear{2025}
\acmYear{2025}
\setcopyright{acmlicensed}
\acmConference[SIGGRAPH Conference Papers '25]{Special Interest Group on Computer Graphics and Interactive Techniques Conference Conference Papers }{August 10--14, 2025}{Vancouver, BC, Canada}
\acmBooktitle{Special Interest Group on Computer Graphics and Interactive Techniques Conference Conference Papers (SIGGRAPH Conference Papers '25), August 10--14, 2025, Vancouver, BC, Canada}
\acmDOI{10.1145/3721238.3730605}
\acmISBN{979-8-4007-1540-2/2025/08}

\title{HumanRAM: Feed-forward Human Reconstruction and Animation Model using Transformers}

\author{Zhiyuan Yu}
\orcid{0000-0001-7220-7789}
\email{zyuaq@ust.hk}
\affiliation{
 \institution{Hong Kong University of Science and Technology}
 \department{Mathematics}
 \city{Hong Kong}
 \country{China}
}
\authornote{The first two authors contributed equally to this work.}
\authornote{Work done during an internship at Huawei.}

\author{Zhe Li}
\orcid{0000-0003-4703-0875}
\email{lizhe_thu@126.com}
\affiliation{
 \institution{Huawei}
 \city{Hangzhou}
 \country{China}
}
\authornotemark[1]

\author{Hujun Bao}
\orcid{0000-0002-2662-0334}
\email{bao@cad.zju.edu.cn}
\affiliation{
  \institution{State Key Laboratory of CAD\&CG, Zhejiang University}
  \city{Hangzhou}
  \country{China}
}

\author{Can Yang}
\orcid{0000-0002-4407-3055}
\email{macyang@ust.hk}
\affiliation{
 \institution{Hong Kong University of Science and Technology}
 \department{Mathematics}
 \city{Hong Kong}
 \country{China}
}
\authornote{Corresponding authors.}

\author{Xiaowei Zhou}
\orcid{0000-0003-1926-5597}
\email{xwzhou@zju.edu.cn}
\affiliation{
 \institution{State Key Laboratory of CAD\&CG, Zhejiang University}
 \city{Hangzhou}
 \country{China}
}
\authornotemark[3]
\renewcommand\shortauthors{Yu, Z. et al}

\begin{document}

\begin{abstract}
3D human reconstruction and animation are long-standing topics in computer graphics and vision. 
However, existing methods typically rely on sophisticated dense-view capture and/or time-consuming per-subject optimization procedures. 
To address these limitations, we propose HumanRAM, a novel feed-forward approach for generalizable human reconstruction and animation from monocular or sparse human images.
Our approach integrates human reconstruction and animation into a unified framework by introducing explicit pose conditions, parameterized by a shared SMPL-X neural texture, into transformer-based large reconstruction models (LRM).
Given monocular or sparse input images with associated camera parameters and SMPL-X poses, our model employs scalable transformers and a DPT-based decoder to synthesize realistic human renderings under novel viewpoints and novel poses. 
By leveraging the explicit pose conditions, our model simultaneously enables high-quality human reconstruction and high-fidelity pose-controlled animation.
Experiments show that HumanRAM significantly surpasses previous methods in terms of reconstruction accuracy, animation fidelity, and generalization performance on real-world datasets.
Video results are available at \url{https://zju3dv.github.io/humanram/}.
\end{abstract}

%
%


%
%


\begin{teaserfigure}
    \includegraphics[width=\textwidth]{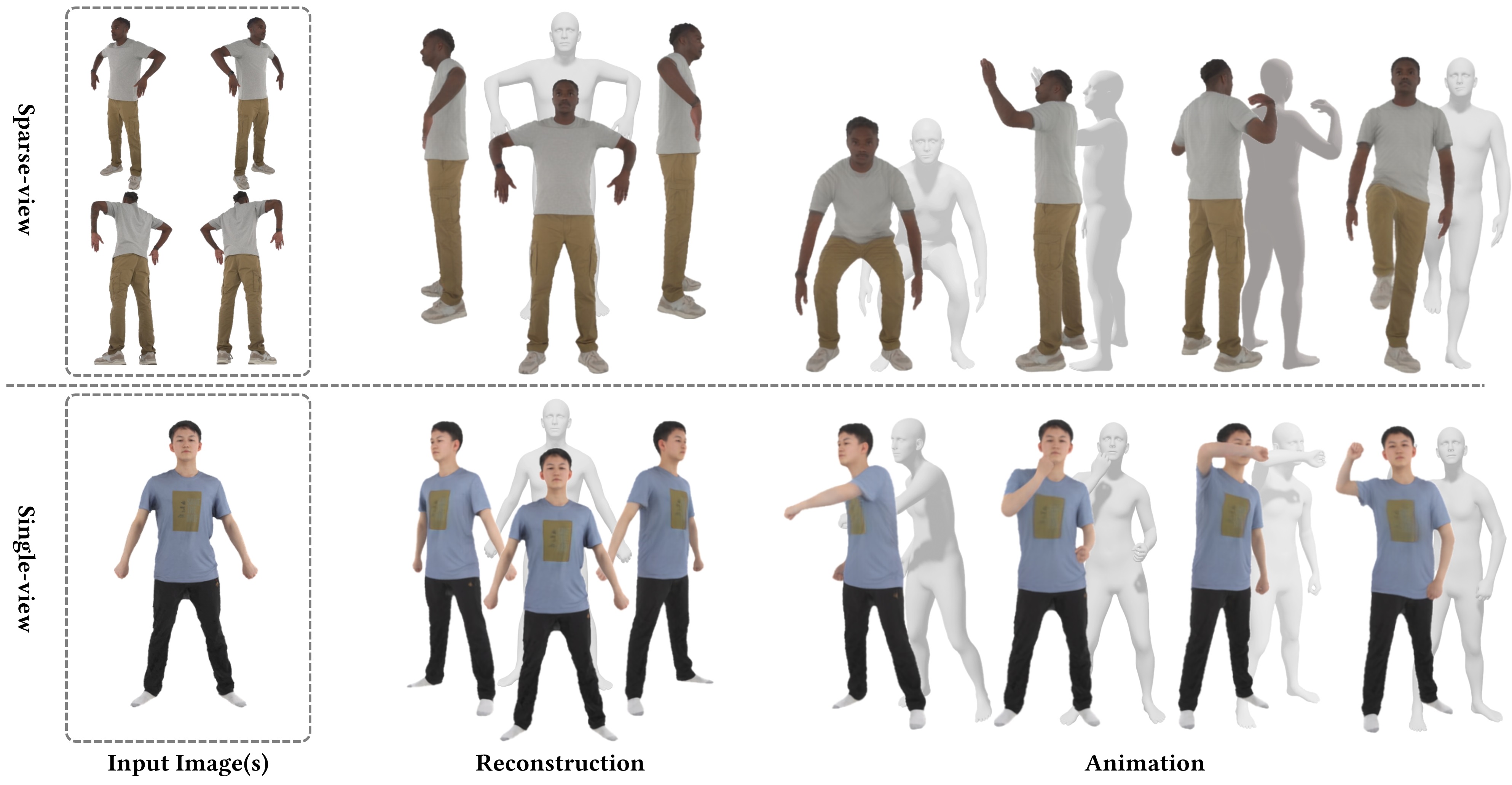}
    \caption{
    \textbf{We propose HumanRAM, a novel approach for feed-forward novel view synthesis (reconstruction) and novel pose synthesis (animation) from sparse/single-view human image(s). The animation poses are from ActorsHQ~\cite{isik2023humanrf} and AMASS~\cite{AMASS}.}
    }
    \Description{Teaser image showing results of HumanRAM for human reconstruction and animation.}
    \label{fig:teaser}
\end{teaserfigure}

\maketitle

\section{INTRODUCTION}
Reconstruction and animation are two core topics in human-centric 3D vision and graphics.
Although high-end dense-view capture systems and modeling technologies~\cite{collet2015high,guo2019relightables,isik2023humanrf,li2024animatablegaussians} achieve high-quality 3D human reconstruction and animation, the complicated hardware and time-consuming per-subject optimization limit their broader applications.

This bottleneck leads to a trend of sparse/single-view reconstruction, which employs generalizable feed-forward networks to predict 3D humans directly from limited inputs.
Pioneering works like PIFu~\cite{saito2019pifu,saito2020pifuhd} proposed to reconstruct human geometry via occupancy fields, but paid less attention to photo-realistic rendering demanded by real applications.
Recently, with significant advances in differentiable rendering \cite{mildenhall2020nerf,kerbl20233d,Laine2020diffrast} and neural rendering \cite{thies2019deferred,tewari2020state},
researchers start using scalable transformers~\cite{attention} to predict 3D representations or novel-view renderings from input images.
These approaches originate from the Large Reconstruction Model (LRM) \cite{hong2023lrm}.
After being exposed to a large amount of 3D data (e.g., Objaverse~\cite{deitke2023objaverse}), LRM and its follow-ups \cite{gslrm2024,wei2024meshlrm,jin2024lvsm} learn to predict 3D models or images in a single forward pass.
While LRMs achieve feed-forward 3D reconstruction, they struggle with fine-grained details in human geometry under complex poses and cloth deformations.
Moreover, existing frameworks focus on static reconstruction, ignoring dynamic animations that are essential for interactive applications.

To address these challenges, we propose Human Reconstruction and Animation Model (HumanRAM), a novel framework that integrates human reconstruction and animation into a unified feed-forward model.
We leverage Large View Synthesis Model (LVSM) \cite{jin2024lvsm} as the foundational architecture, which implicitly learns 3D structures and directly regresses novel-view renderings. Previous explicit 3D representations usually require precise geometry for high-quality output. However, geometric constraints are insufficient under sparse observations. By harnessing the implicit nature of LVSM, we overcome this limitation and improve the model's performance and generalization capacity.

Original LVSM maps input images \& cameras, as well as the target camera, into patch tokens and regresses the target-view image using transformers.
To endow LVSM with the animation ability and improve the reconstruction quality for humans, we introduce SMPL-X \cite{SMPL-X:2019}, a parametric human mesh model that provides strong pose and geometry priors as additional input tokens.
Given calibrated multi-view human images, SMPL-X can be estimated using off-the-shelf tools~\cite{sun2024aios,zhang2021lightweight}.
To tokenize the SMPL-X prior, we introduce rasterization of a shared neural texture map bound to the SMPL-X mesh across input and target views. 
This process yields pose images that spatially align with the RGB and camera tokens used in LVSM.
These pose images serve as a strong geometrical and semantical guide for the transformer's attention mechanism, thereby enabling a more realistic novel-view synthesis and pose-controlled animation.

The key insights of our method are: 1) The rasterized pose images establish shared embedding space between input and target views, providing explicit correspondences for the self-attention layers of transformers to reassemble the target view, thus producing higher-fidelity reconstruction.
2) The pose images enable LVSM to match appearance across diverse poses through the shared neural texture map, thereby achieving realistic animation.
Moreover, we propose a DPT-based~\cite{Ranftl2021dpt} decoder to facilitate the information exchange among neighboring patches and intermediate transformer features, effectively suppressing checkerboard artifacts prevalent in linear decoders.

Overall, the synergy of design choices enables high-quality human reconstruction and photo-realistic human animation from sparse/single image(s), as shown in Fig.~\ref{fig:teaser} and Fig.~\ref{fig:results}.
\section{RELATED WORK}
\subsection{Generalizable Human Reconstruction}
Human reconstruction has been widely explored over the past few decades.
Traditional methods reconstruct human geometry and texture from dense-view images~\cite{starck2007surface,liu2009point,wu2011shading,vlasic2009dynamic,bradley2008markerless,collet2015high,guo2019relightables}.
With advancements in differentiable 3D representations like implicit functions~\cite{mescheder2019occupancy,peng2020convolutional,park2019deepsdf,chabra2020deep}, neural radiance fields (NeRF) \cite{mildenhall2020nerf}, and 3D Gaussian splatting (3DGS) \cite{kerbl20233d}, researchers tend to learn data-driven feed-forward models.
Methods like BodyNet \cite{varol2018bodynet}, DeepHuman \cite{zheng2019deephuman}, and \citet{tang2023high} regress volumetric outputs from image(s) but face resolution limits from GPU memory constraints. 
PIFu~\cite{saito2019pifu} and its successors~\cite{zheng2021pamir,xiu2022icon,xiu2023econ,saito2020pifuhd,yu2021function4d,zheng2021deepmulticap,zhang2024gta,zhang2024sifu,Yang_2024_CVPR_HiLo,cao2023sesdf} address this by learning pixel-aligned implicit functions for human geometry recovery.

In recent years, many works have developed generalizable models based on NeRF or 3DGS for human novel view synthesis.
Similar to PIFu, NeRF-based approaches~\cite{raj2021pva,kwon2021neural,shao2022doublefield,hu2023sherf,mihajlovic2022keypointnerf,HDhuman2024tvcg,lin2022efficient,chen2024dihur,sun2024metacap,chen2023gmnerf} extract pixel-aligned features and learn image-conditioned radiance fields.
In contrast, 3DGS-based methods explicitly parameterize 3D gaussians in pixel space~\cite{zheng2024gps,zhou2024gps+,tu2024tele,hu2024eva,dong2024gaussian}, UV space~\cite{kwon2024ghg} or tokens~\cite{gst}.
More recently, 3D AIGC has made remarkable progress~\cite{dreamfusion,liu2023zero,liu2024one,voleti2024sv3d}, leading researchers to model the reconstruction as an image-conditioned generation task~\cite{jnchen24hgm,albahar2023single,humanlrm2023,SiTH,havefun,PuzzleAvatar,DiffHuman,Diffusion-FOF,ConTex-Human,liu2024humangaussian,huang2024humannorm,xu2023seeavatar,cao2024dreamavatar,kolotouros2024dreamhuman,he2024magicman}.
Although diffusion models enhance texture hallucination for occluded regions, their iterative refinement process incurs higher computational costs than feed-forward models.

\subsection{Large Reconstruction Model}
Large Reconstruction Model (LRM) was first proposed by ~\citet{hong2023lrm}, which learns a generalizable NeRF~\cite{mildenhall2020nerf} from a single image. Subsequent  works~\cite{wang2023pf,xie2024lrmzero,humanlrm2023,xu2023dmv3d,wei2024meshlrm,tang2025lgm,xu2024grm,gslrm2024} explore LRM in various downstream tasks.
For instance, PF-LRM~\cite{wang2023pf} learns from unposed images. 
LRM-Zero~\cite{xie2024lrmzero} and MegaSynth~\cite{jiang2024megasynth} train LRM on synthetic data and successfully generalize to real data. 
DMV3D~\cite{xu2023dmv3d} applies LRM as a diffusion denoiser to improve generation view consistency.
Some researchers extend the representation of LRM from NeRF to mesh~\cite{wei2024meshlrm}, 3DGS~\cite{tang2025lgm,xu2024grm,gslrm2024,liang2024bullet,shen2024gamba,yi2024mvgamba,ziwen2024llrm} and 2DGS~\cite{LaRa}. 
More recently, LVSM~\cite{jin2024lvsm} synthesizes novel views using pure transformers. 
Despite these advances, existing LRM variants mainly focus on object/scene reconstruction, 
ignoring human-centric applications. 
In contrast, our method specializes in human reconstruction and animation.

\subsection{Human Animation}
Human animation aims to generate novel-pose images given one or more input images. 
Previous works are categorized into 2D and 3D animation. 
2D animation formulates the task as signal-driven image generation~\cite{chan2019dance,lwb2019,Zhang_2022_CVPR,zhao2022thin,siarohin2021motion,Siarohin_2019_CVPR,Siarohin_2019_NeurIPS,yu2023bidirectionally,ren2020deep}.
Recently, diffusion-based methods~\cite{ma2023followpose,hu2023animateanyone,wang2023disco,mimicmotion2024,zhu2024champ,xu2024magicanimate,men2024mimo,shao2024human4dit} have gained huge attention for their powerful generation capabilities, but they suffer from time-consuming generation due to the denoising process.

3D methods typically optimize person-specific avatars from single or multi-view videos using various 3D representations (e.g., point clouds~\cite{su2023npc}, mesh~\cite{bagautdinov2021driving,chen2024meshavatar}, implicit field~\cite{peng2024animatable,ARAH_2022_ECCV,jiang2022selfrecon,xu2024relightable,li2023posevocab,zhang2023explicifying}, NeRF~\cite{peng2021animatable,liu2021neural,jiang2022instantavatar,jiang2022neuman,yu2023monohuman,NECA2024CVPR,xu2022sanerf,li2022tava,liu2024texvocab} and 3DGS~\cite{kocabas2024hugs,lei2024gart,li2024animatablegaussians,moon2024exavatar,shao2024splattingavatar,wen2024gomavatar,hu2024expressive,zielonka25dega,lin2024layga}).
The avatars are then animated using linear blend skinning (LBS).
However, the optimization process is time-intensive and can fail with very sparse inputs.
To generalize, researchers use learned priors~\cite{mu2023actorsnerf,chatziagapi2024migs,ho2023custom} or feed-forward models~\cite{MPS-NeRF,gao2023neural,kwon2023neural,he2021arch++,huang2020arch,shin2025canonicalfusion}.
Our method bridges human prior and Large Reconstruction Model, leading to more realistic animation.
\begin{figure*}[t]
    \centering
    \includegraphics[width=1.0\textwidth]{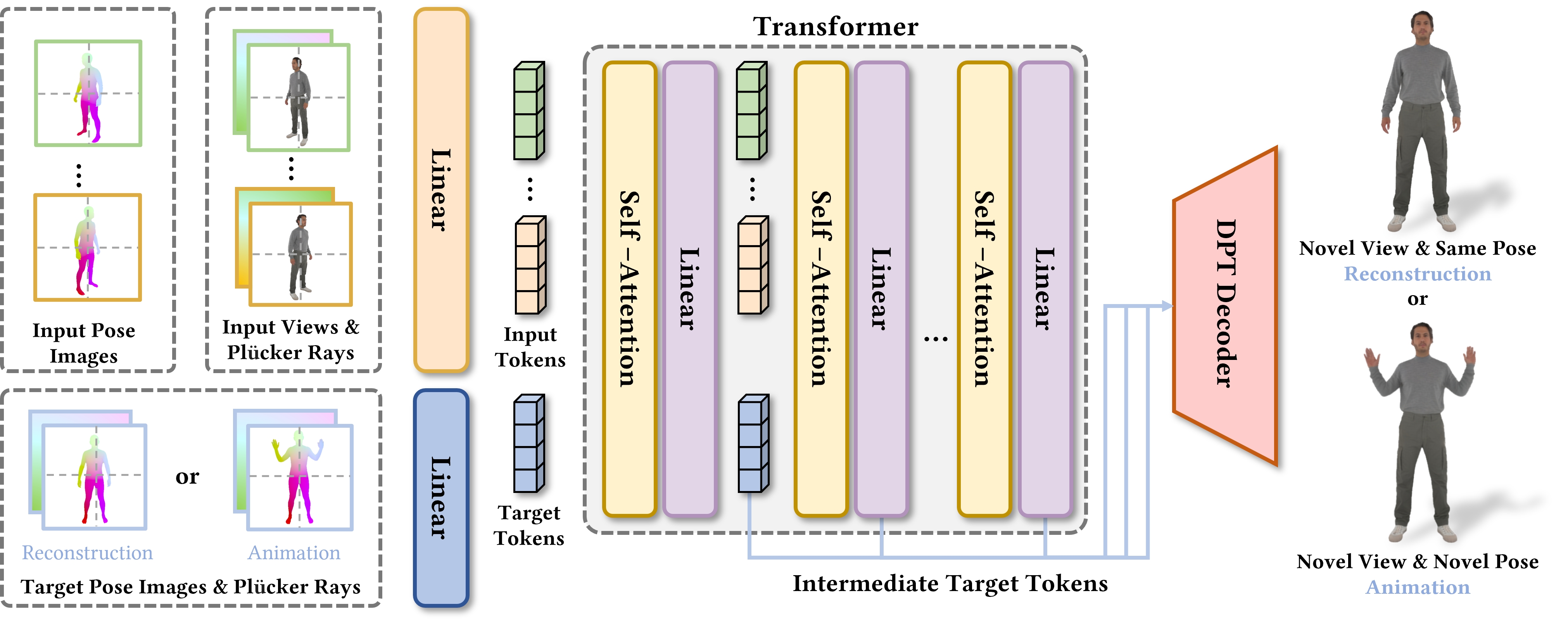}
    \caption{
    \textbf{Pipeline of HumanRAM.} HumanRAM adopts transformers for human reconstruction and animation from sparse view images in a feed-forward manner.
    We first patchify and project spare-view RGB images and their corresponding Pl\"ucker rays and pose images into input tokens through a linear layer.
    The pose images are acquired by rasterizing the SMPL-X neural texture onto the input views.
    Similarly, given the target novel view under the same or another novel pose, the target tokens are created from the target Pl\"ucker rays and pose images through another linear layer. 
    Then both input tokens and target tokens are fed into transformer blocks. 
    Finally, a DPT-based decoder regresses the intermediate target tokens to a high-fidelity human image under the target view and target pose.
    Overall, HumanRAM realizes feed-forward reconstruction and animation by controlling the target views and target poses at the input end.
    }
    \label{fig:overview}
\end{figure*}

\section{METHOD}\label{method}
\subsection{Preliminary: LVSM}
Large View Synthesis Model (LVSM)~\cite{jin2024lvsm} is a recent method for neural rendering without using any explicit 3D representations.
This method inputs multi-view images and camera parameters and outputs target-view renderings through encoder-decoder or decoder-only transformers. 
Specifically, given $N$ images with their corresponding camera poses parameterized by Pl\"ucker ray embeddings \cite{plucker}, denoted as
$\{\mathbf{I}_i\in\mathbb{R}^{H\times W\times3},\mathbf{P}_i\in\mathbb{R}^{H\times W\times6}|i=1,...,N\}$, LVSM first maps them into patch tokens $\mathbf{x}_{ij}\in\mathbb{R}^d$ with a linear layer ($d$ is the token dimension):
\begin{equation}\label{eq:1}
    \mathbf{x}_{ij} = \text{Linear}_{\text{inp}}([\mathbf{I}_{ij},\mathbf{P}_{ij}]),
\end{equation}
where $\mathbf{I}_{ij}\in\mathbb{R}^{3p^2}$ and $\mathbf{P}_{ij}\in\mathbb{R}^{6p^2}$ mean the $j$-th $p\times p$ patch of $\mathbf{I}_i$ and $\mathbf{P}_i$, and $[\cdot,\cdot]$ means concatenation.
The target-view pose is also represented as Pl\"ucker ray embedding $\mathbf{P}^t\in\mathbb{R}^{H\times W\times6}$ and mapped to patch tokens $\mathbf{q}_j\in\mathbb{R}^d$ with another linear layer:
\begin{equation}\label{eq:2}
    \mathbf{q}_j = \text{Linear}_{\text{tar}}(\mathbf{P}^t_{j}).
\end{equation}

Given input and target tokens, decoder-only LVSM synthesizes target-view tokens $\mathbf{y}_j\in\mathbb{R}^d$ through transformers $\mathcal{T}$:
\begin{equation}
    \mathbf{x}'_i,...,\mathbf{x}'_{l_x},\mathbf{y}_1,...,\mathbf{y}_{l_q}=\mathcal{T}(\mathbf{x}_i,...,\mathbf{x}_{l_x},\mathbf{q}_1,...,\mathbf{q}_{l_q}),
\end{equation}
where $l_x$ and $l_q$ mean the number of input and target tokens.
Finally, LVSM regresses the RGB values of each target patch from output tokens with a linear layer followed by a Sigmoid function:
\begin{equation}\label{eq:4}
    \hat{\mathbf{I}}_j^t=\text{Sigmoid}(\text{Linear}_{\text{out}}(\mathbf{y}_j))\in\mathbb{R}^{{3p^2}}.
\end{equation}
The predicted RGB values are unpatchfied to 2D space to form the final target image. 
In this paper, we incorporate dedicated designs into LVSM for human reconstruction and animation.

\subsection{Overview}
Given sparse-view images of a character, we aim to synthesize the character under novel views and poses, i.e., to perform feed-forward human reconstruction and animation. 
As a state-of-the-art feed-forward large reconstruction model (LRM), LVSM~\cite{jin2024lvsm} is introduced as a foundational architecture of our method.
To endow LRM with the animation ability, we introduce pose tokens parameterized by a neural texture \cite{thies2019deferred} bound with SMPL-X~\cite{SMPL-X:2019} into LVSM.
Specifically, as illustrated in Fig.~\ref{fig:overview}, we render the learnable SMPL-X neural texture to the sparse input views, resulting in $N$ feature maps, which we refer to as pose images.
The input RGB images, their corresponding Pl\"ucker embeddings, and the pose images are concatenated and then patchfied as \textit{input tokens}.
Given the target view and the target human poses to be synthesized, we similarly concatenate and patchify them as \textit{target tokens}.
The input and target tokens are fed into a transformer model, and the \textit{output tokens} are regressed to produce the synthesized human image under the target view and pose.

\subsection{Pose-conditioned Reconstruction and Animation}\label{sec:3.3}
Since LVSM discards explicit 3D representation, we cannot directly use the SMPL-X~\cite{SMPL-X:2019} model as a geometry prior or proxy for pose-conditioned reconstruction and animation, as done in previous works \cite{xiu2023econ,zheng2021pamir,kwon2024ghg,huang2020arch,taubner2024cap4d}.
Inspired by neural texture~\cite{thies2019deferred,deng2024ram,yoon2022learning}, we render the SMPL-X mesh with a neural texture onto multi-view 2D image planes, generating multi-view pose conditions. 
These pose conditions serve not only as a strong geometry prior for the novel view synthesis but also as an enabler for the animation ability.

\begin{figure}[!htbp]
    \centering
    \includegraphics[width=1.0\linewidth]{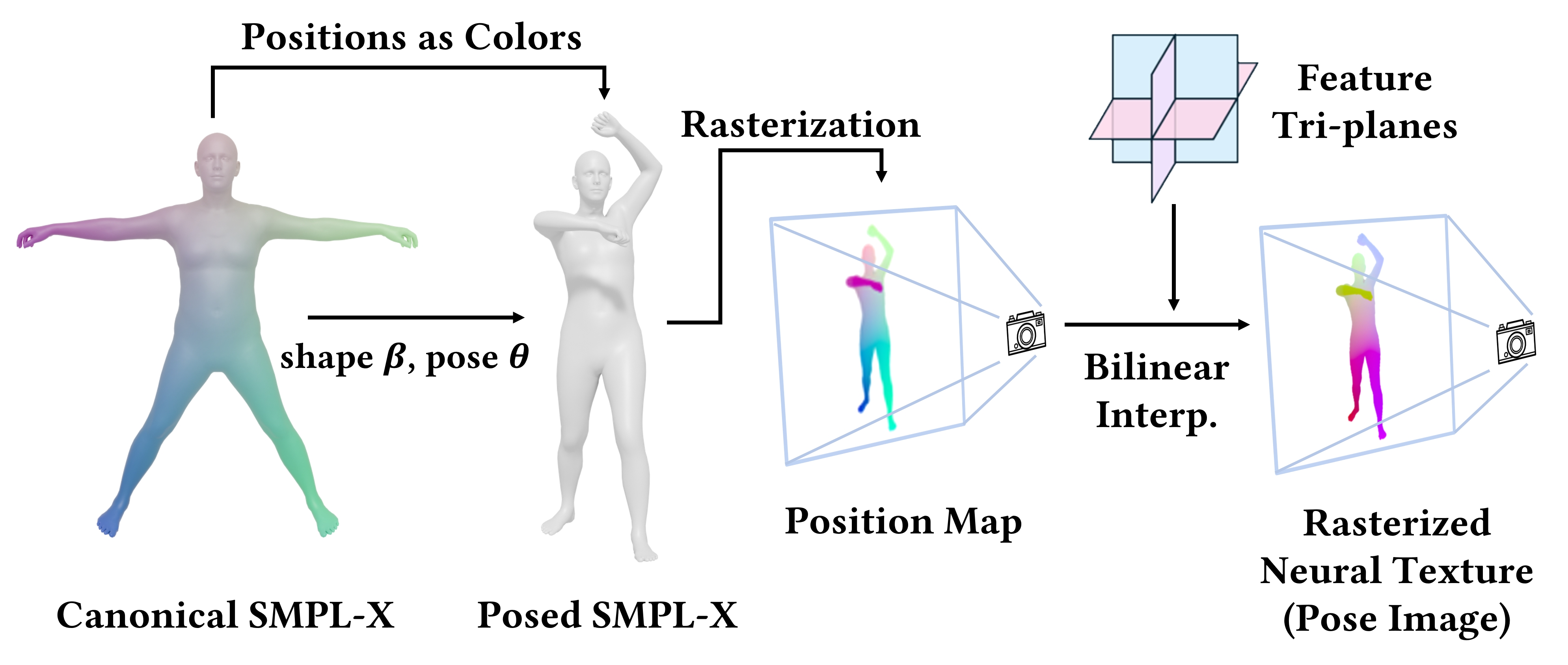}
    \caption{
    \textbf{Illustration of the process of neural texture rasterization.} 
    We first render position maps with canonical SMPL-X as vertex colors, and then the position maps are used to sample triplane-based neural texture.
    }
    \label{fig:neutex}
\end{figure}

\paragraph{SMPL-X Neural Texture}
We adopt tri-planes \cite{Chan2021eg3d} to represent the neural texture for its effectiveness and compactness.
As illustrated in Fig.~\ref{fig:neutex}, the neural texture is defined as learnable feature tri-planes within a canonical space, determined by SMPL-X with canonical pose and mean shape.
We denote the canonical SMPL-X vertices as $\mathbf{V}_\text{cano}\in\mathbb{R}^{N_{V}\times3}$ and feature tri-planes as $\mathbf{T}\in\mathbb{R}^{3\times H'\times W'\times C}$, where $H'$ and $W'$ are the resolution of each plane, and $C$ is the feature dimension.
For each position $\mathbf{v}\in\mathbb{R}^3$ on the canonical SMPL-X surface, its corresponding neural texture is the concatenation of sampled features on each plane:
\begin{equation}
\label{eq: neural texture}
\begin{aligned}
    \mathbf{F}(\mathbf{v};\mathbf{T})
    &=[\text{BLerp}(\mathbf{v}^{xy};\mathbf{T}^{xy}),\\&\text{BLerp}(\mathbf{v}^{xz};\mathbf{T}^{xz}),\text{BLerp}(\mathbf{v}^{yz};\mathbf{T}^{yz})] \in\mathbb{R}^{3C},
\end{aligned}
\end{equation}
where $\text{BLerp}(\cdot)$ is the bilinear interpolation function on the feature plane given 2D query coordinates.
The SMPL-X neural texture is shared across all the identities, providing guidance for pose-conditioned reconstruction and animation.

\paragraph{Input Tokens}
Given calibrated multi-view input images of a character, we obtain the registered SMPL-X mesh using multi-view motion capture like \cite{zhang2021lightweight}.
The registered SMPL-X serves as a geometry proxy. We then bind its vertex attributes with the canonical positions $\mathbf{V}_\text{cano}$ and rasterize it to the input views, producing $N$ position maps \{$\mathbf{V}_i\in\mathbb{R}^{H\times W\times3}$\}, where each pixel corresponds to a canonical position.
The rendered position maps are then used to sample the neural texture using Eq.~\ref{eq: neural texture}. 
Consequently, we rasterize the SMPL-X neural texture onto the input views, obtaining $N$ pose images $\{\mathbf{F}_i\in\mathbb{R}^{H\times W\times 3C}\}$.
Then, similar to Eq.~\ref{eq:1}, we concatenate RGB images, Pl\"ucker embeddings, and pose images along the channel dimension, and then patchify them as ``input tokens'':
\begin{equation}
\label{eq: input tokens}
    \mathbf{x}_{ij} = \text{Linear}_{\text{inp}}([\mathbf{I}_{ij},\mathbf{P}_{ij}, \mathbf{F}_{ij}]).
\end{equation}

\paragraph{Target Tokens for Reconstruction}
Given a new target viewpoint, we rasterize the neural texture onto it using the registered SMPL-X model to obtain a novel view pose image $\mathbf{F}^{t}$.
Similar to Eq.~\ref{eq:2}, we concatenate the Pl\"ucker embeddings and pose image of the target view, and then patchify them as ``target tokens'' for reconstruction:
\begin{equation}
\label{eq: target tokens}
    \mathbf{q}_j^{\text{recon}} = \text{Linear}_{\text{tar}}([\mathbf{P}^t_{j},\mathbf{F}^t_{j}]).
\end{equation}

\paragraph{Target Tokens for Animation}
Since the rasterized neural texture provides rich human pose information, it is natural to explore our model's potential animation ability for both novel view and pose synthesis. 
Specifically, given a novel target pose $\hat{\theta}$, we first transform it into a posed SMPL-X model.
We then rasterize the neural texture onto a novel view using the posed SMPL-X, obtaining the pose image $\hat{\mathbf{F}}^{t}$ representing both the novel view and the novel pose.
Following Eq.~\ref{eq: target tokens}, we acquire ``target tokens'' $\{\mathbf{q}_j^{\text{ani}}\}$ for animation.

\paragraph{Output Tokens}
The input and target tokens are subsequently fed to a decoder-only transformer $\mathcal{T}$, composed of a series of self-attention layers, producing a sequence of ``output tokens''.
These output tokens and intermediate features are decoded as an image under the target view and target pose using a DPT-based decoder \cite{Ranftl2021dpt} (Sec.~\ref{subsec: dpt-based unsampler}).

\paragraph{Discussion on Pose Conditions}
We discuss the impact of the neural texture-based pose images on both reconstruction and animation.
\begin{itemize}
    \item From the perspective of novel view synthesis, i.e., reconstruction, LVSM \cite{jin2024lvsm} learns cross-view matching from RGB and camera pose information to reassemble a novel view image using the attention mechanism.
    Our method introduces additional SMPL-X neural texture into the matching process, providing more explicit correspondences for higher-quality view synthesis, as demonstrated in Fig.~\ref{fig:thuman2.1_recon}, compared to vanilla LVSM.
    \item On the other end of the spectrum, the pose condition enables texture matching across different poses with a shared neural texture, achieving novel pose synthesis.
    To the best of our knowledge, HumanRAM is the first to endow Large Reconstruction Model with the animation ability.
\end{itemize}

\subsection{DPT-based Decoder}\label{subsec: dpt-based unsampler}
Original LVSM~\cite{jin2024lvsm} uses a linear layer to decode tokens into RGB values directly. 
We empirically find that such a simple decoder yields patch-like artifacts for humans, especially in regions suffering severe self-occlusions or containing thin structures, as shown in Fig. ~\ref{fig:ablation}. 
We hypothesize that such artifacts are attributed to the lack of information exchange between neighboring patch tokens when decoding. 
Inspired by the dense prediction transformers (DPT) used in various vision transformer models~\cite{oquab2023dinov2,Ranftl2021dpt,wang2024moge,depth_anything_v2}, we replace the linear layer with stacks of residual CNN layers, similar to DPT heads, enhancing the local information fusion. 
Therefore, the final synthesized image $\hat{\mathbf{I}}^t$ is formed using a DPT-based decoder:
\begin{equation}
    \hat{\mathbf{I}}^t=\text{Sigmoid}(\text{DPT}(\{\mathbf{y}^i|i=3,6,9, 12\}),
\end{equation}
where $\mathbf{y}^i$ denotes the intermediate tokens of the $i$-th layer.

\subsection{Loss Functions}\label{loss}
Given the predicted target-view images $\{\mathbf{\hat{I}}_i\in\mathbb{R}^{H\times W\times3}|i=1,...,M\}$, we optimize HumanRAM using the following objective:
\begin{equation}
    \mathcal{L}=\dfrac{1}{M}\Sigma_{i=1}^{M}(\mathcal{L}_{\mathrm{MSE}}(\mathbf{\hat{I}}_i,\mathbf{I}_i)+\lambda\cdot\mathcal{L}_{\mathrm{Perc}}(\mathbf{\hat{I}}_i,\mathbf{I}_i)),
\end{equation}
where $\mathcal{L}_{MSE}$ denotes the mean squared error
and $\mathcal{L}_{\text{Perc}}$ denotes the perceptual loss~\cite{chen2017photographic}, computing $L_1$ difference between the extracted features from the VGG-19 network $\Phi$~\cite{simonyan2014very}.
$\lambda$ is the loss weight of $\mathcal{L}_{\text{Perc}}$ and set to 1.0 in our experiments.
\begin{figure*}[!htbp]
    \centering
    \includegraphics[width=1.0\textwidth]{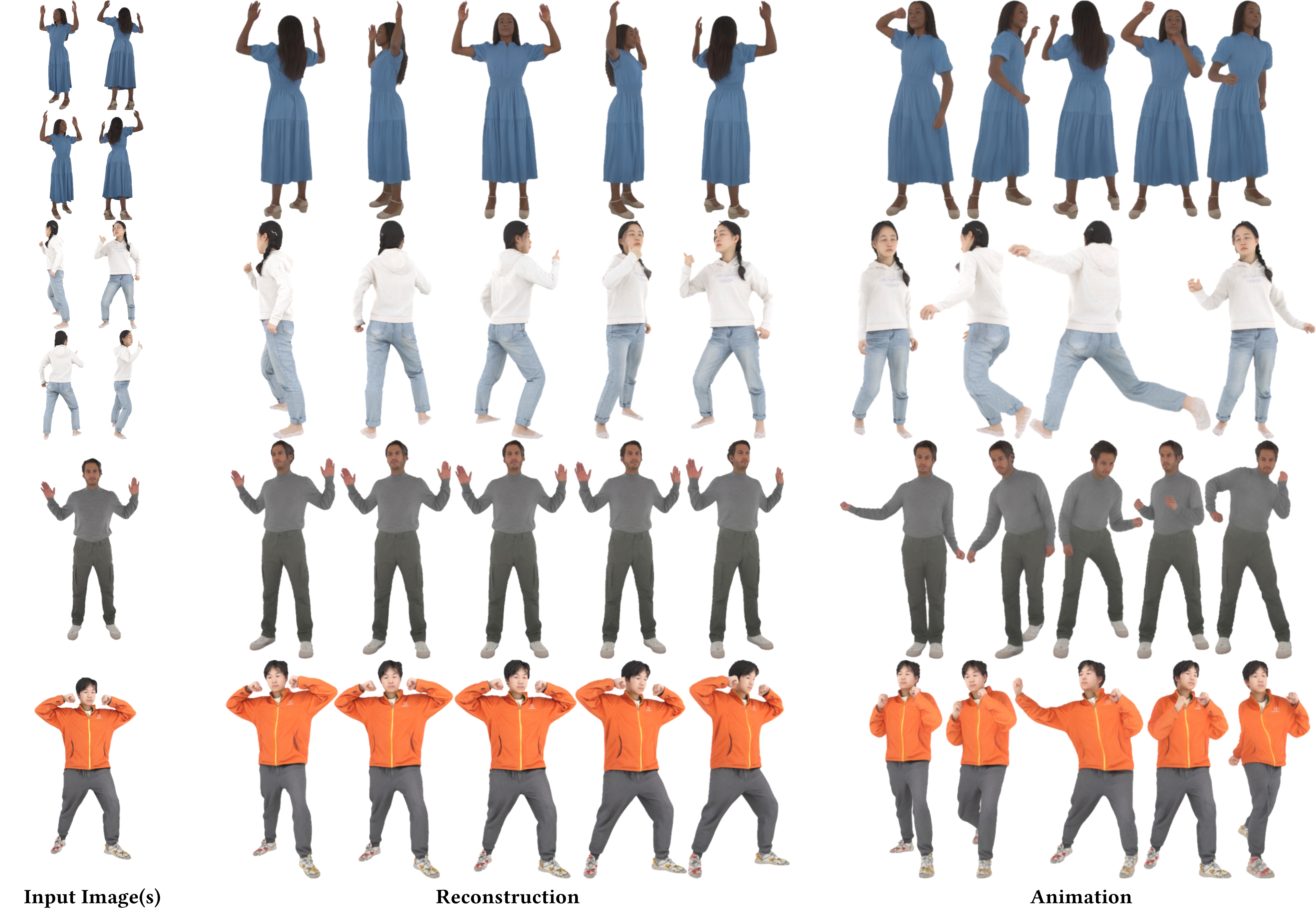}
    \caption{
    \textbf{Qualitative results on ActorsHQ~\cite{isik2023humanrf} and THuman2.1~\cite{yu2021function4d}.} The top two rows show the reconstruction and animation results from multi-view inputs, while the bottom two rows show the results from single-view input. The driving poses for animation are from ActorsHQ \cite{isik2023humanrf} and AMASS~\cite{AMASS}.
    }
    \label{fig:results}
\end{figure*}

\section{EXPERIMENTS}\label{exp}
As shown in Fig.~\ref{fig:teaser} and Fig.~\ref{fig:results}, our method can create realistic human reconstruction and animation from single and sparse images since our transformer-based architecture is flexible to the image token number.
Video results can be found in the Supp. video.

\subsection{Settings}\label{settings}
The implementation and training details are presented in the Supp. document.
\paragraph{Datasets.}
We conduct experiments on four public datasets: \textit{THuman2.1}~\cite{yu2021function4d},
\textit{Human4DiT}~\cite{shao2024human4dit},
\textit{ZJUMoCap}~\cite{peng2021neural}, and 
\textit{ActorsHQ}~\cite{isik2023humanrf} for training and evaluation.
THuman2.1 and Human4DiT 
comprise thousands of high-quality 3D human scans, texture maps, and SMPL-X~\cite{SMPL-X:2019} fittings.
We use the training set of 2300 scans from THuman2.1. 
The training scans are categorized according to human identities, enabling the model to learn animation across different poses of the same identity.
Each training scan is normalized into a $[-1, 1]^3$ bounding box and rendered to 60-view images at a resolution of 512 via Cycles~\cite{blender}.
The cameras are randomly sampled with an altitude of $[-45^{\circ},45^{\circ}]$ and a radius of $[2.0, 3.0]$.
ZJUMoCap and ActorsHQ are human avatar datasets that provide multi-view human videos and SMPL(-X)s. 
We convert the SMPL~\cite{SMPL:2015} parameters into SMPL-X for ZJUMoCap and use the SMPL-X provided by ~\citet{li2024animatablegaussians} for ActorsHQ.
All images are resized to 512$\times$512 for aligning the input resolution of networks.

\paragraph{Baselines.} 
We compare with generalizable human reconstruction methods GPS-Gaussian~\cite{zheng2024gps} and GHG~\cite{kwon2024ghg}.
We also compare against LRM-like methods LaRa~\cite{LaRa} and LVSM~\cite{jin2024lvsm}. 
For animation, we compare with generalizable human avatar methods NNA~\cite{gao2023neural} and SHERF~\cite{hu2023sherf}, as well as a personalized avatar method 3DGS-Avatar~\cite{qian20233dgsavatar}.

\paragraph{Metrics.} We utilize the Peak Signal-to-Noise Ratio (PSNR), Structure
Similarity Index Measure (SSIM)~\cite{wang2004image}, and Learned Perceptual Image Patch Similarity (LPIPS)~\cite{zhang2018perceptual}
as metrics to assess the results quantitatively and qualitatively. 
PSNR and SSIM are evaluated on mask-cropped images, while LPIPS is 
computed on full-size images.

\begin{table*}[ht]
    \centering        
        \caption{
            \textbf{Quantitative comparison of reconstruction on  THuman2.1~\cite{yu2021function4d} and Human4DiT~\cite{shao2024human4dit}}. We report PSNR, SSIM, and LPIPS to evaluate the reconstruction quality. 
            All methods are trained or finetuned on THuman2.1 for fair comparison.
        }
        \label{tab:scan_dataset}
        \begin{tabular}{c|ccccc|ccccc}
        \toprule
        \multirow{2}{*}{Metrics} & \multicolumn{5}{c|}{THuman2.1}          & \multicolumn{5}{c}{Human4DiT}            \\
                                 & Ours & LVSM & GPS-Gaussian & GHG & LaRa & Ours & LVSM & GPS-Gaussian & GHG & LaRa \\ 
        \midrule
        PSNR$\uparrow$&\textbf{30.34} &28.24  &22.11   &21.88  &23.71   
        &\textbf{26.35}    &25.56   &20.87   &19.47   &22.91      \\
        SSIM$\uparrow$&\textbf{0.9535} &0.9396 &0.9007   &0.8780 &0.8913   &\textbf{0.9373}   &0.9247   &0.8953   &0.8539 &0.8900      \\
        LPIPS$\downarrow$&\textbf{0.0184}  &0.0226 &0.0421   &0.0517 &0.0679   &\textbf{0.0211}   &0.0248   &0.0419   &0.0586 &0.0663      \\ 
        \bottomrule
        \end{tabular}
\end{table*}

\begin{figure*}[!htbp]
    \centering
    \includegraphics[width=0.92\textwidth]{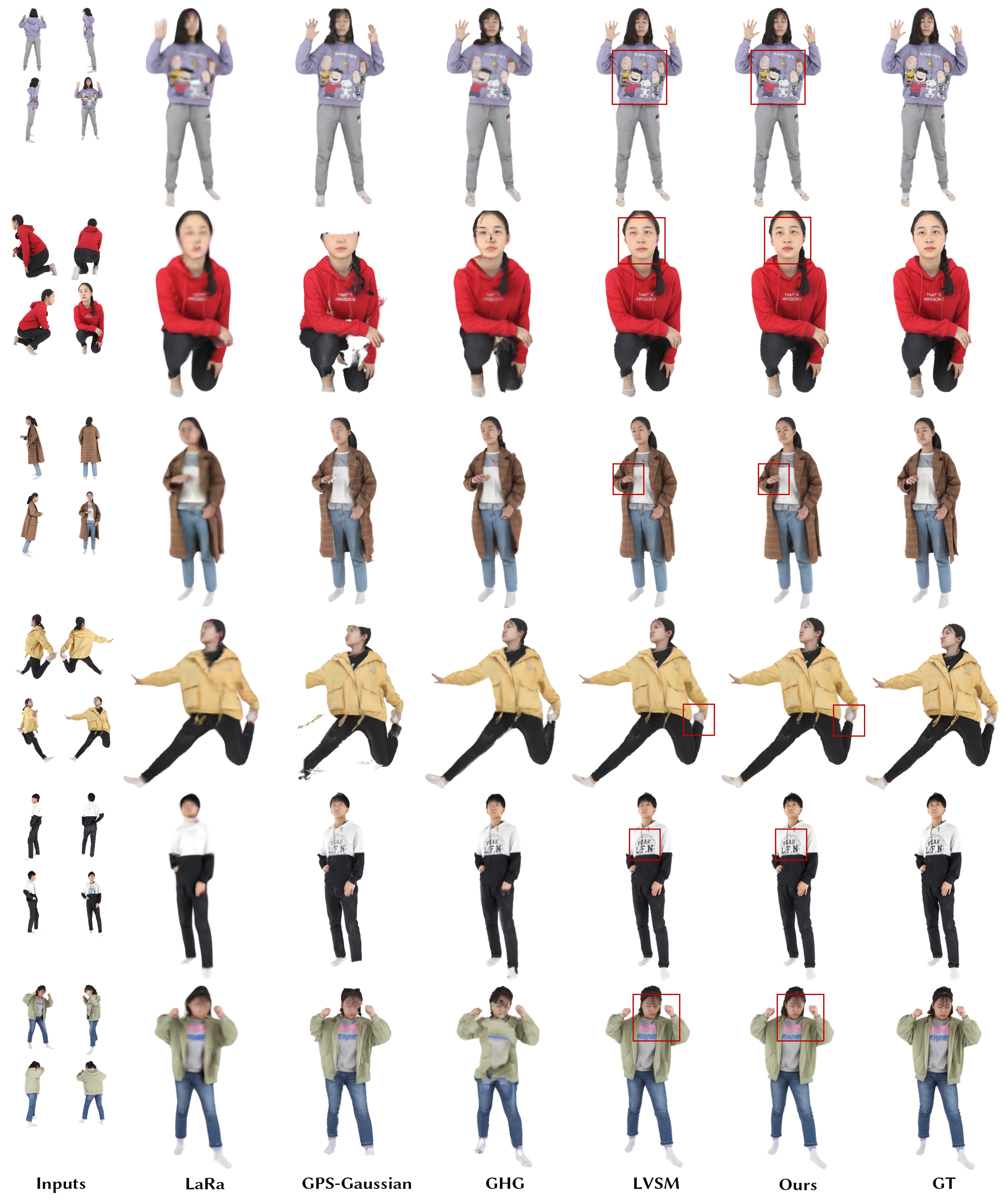}
    \caption{
    \textbf{Qualitative comparisons for reconstruction on THuman2.1~\cite{yu2021function4d} and Human4DiT~\cite{shao2024human4dit}}. We input 4 multi-view images of unseen subjects, and our method achieves a more faithful rendering compared to other reconstruction methods. The first four rows are from THuman2.1 and the last two rows are from Human4DiT. The red boxes indicate the improvements of our method over LVSM~\cite{jin2024lvsm}. 
    }
    \label{fig:thuman2.1_recon}
\end{figure*}

\subsection{Comparison on Reconstruction}
\begin{figure}[ht]
    \centering
    \includegraphics[width=1\linewidth]{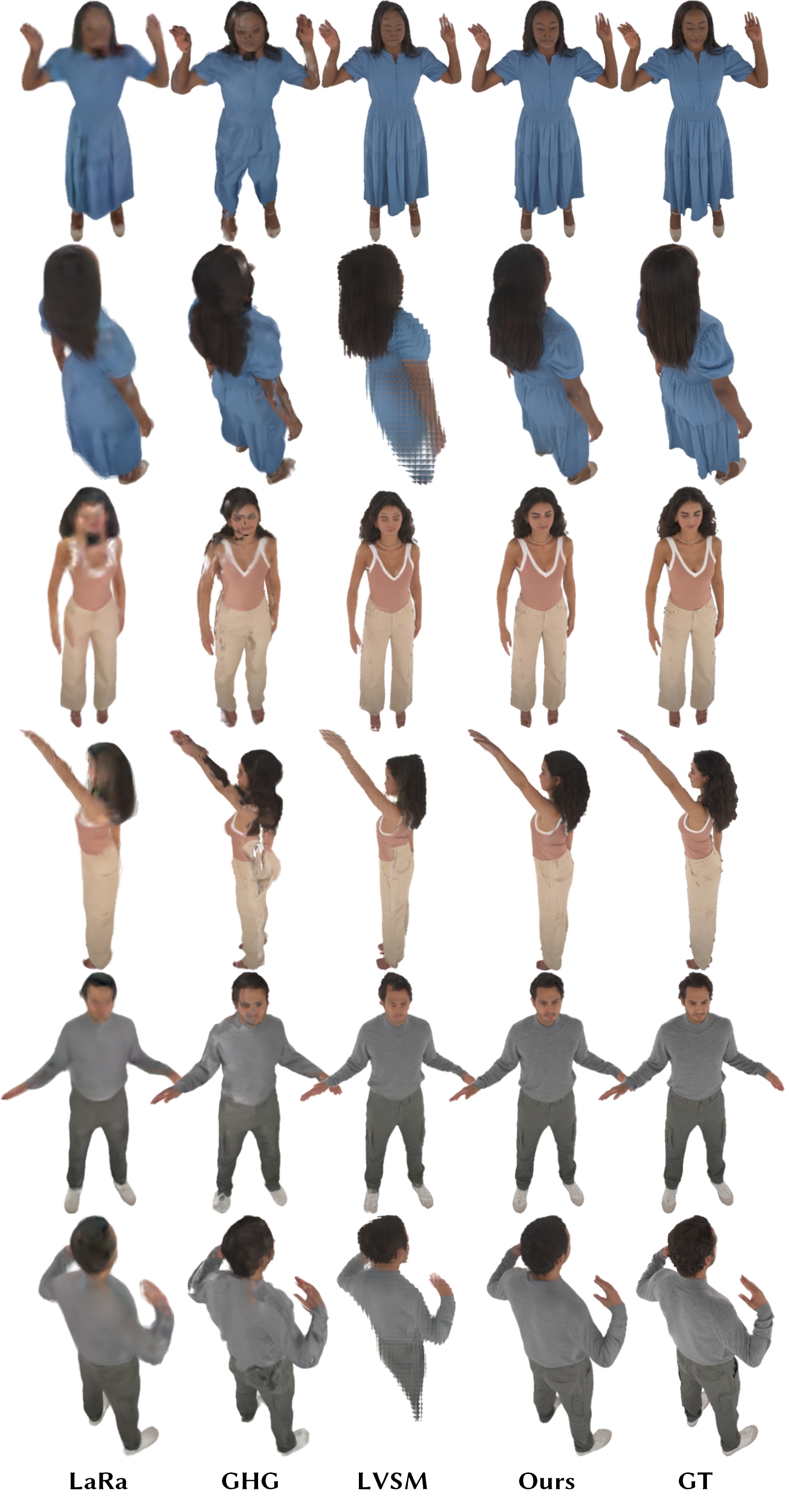}
    \caption{
    \textbf{Qualitative comparisons for reconstruction on ActorsHQ~\cite{isik2023humanrf}}. We input 5 multi-view images, and our method achieves a more faithful rendering compared to other state-of-the-art generalizable reconstruction methods.}
    \label{fig:actorshq_recon}
\end{figure}

We compare HumanRAM with baselines on synthetic and real-world datasets.
For synthetic dataset, we randomly select 200 scans from Thuman2.1~\cite{yu2021function4d} and Human4DiT~\cite{shao2024human4dit} as the test set.
We input 4 uniform views for all methods except GPS-Gaussian~\cite{zheng2024gps}, which requires 5 equal-height images for reasonable stereo rectification.
The qualitative results are shown in Fig. \ref{fig:thuman2.1_recon}.
LaRa~\cite{LaRa} demonstrates blurry results due to its low-resolution volume representation, limiting its ability to model complicated geometries and textures.
GPS-Gaussian is inclined to generate incomplete results because its stereo matching may fail when input views are sparse.
GHG~\cite{kwon2024ghg} applies multi-scaffold SMPL-X~\cite{SMPL-X:2019} mesh as the geometry proxy, which cannot handle loose cloth and tends to produce artifacts if severe self-occlusion occurs.
LVSM~\cite{jin2024lvsm} fails to synthesize fine-grained structures like hands and faces due to the lack of human priors.
Tab. \ref{tab:scan_dataset} reports the numerical comparison on reconstruction.
Overall, our method significantly outperforms previous methods both qualitatively and quantitatively.
To demonstrate the generalization ability of HumanRAM, we conduct experiments on ActorsHQ~\cite{isik2023humanrf} and in-the-wild images.
For ActorsHQ, we select 5 uniform cameras as input and sample 100 frames per subject for evaluation. 
The results are shown in Tab. \ref{tab:real_dataset} and Fig.~\ref{fig:actorshq_recon}. 
All previous methods fail to generate reasonable results on real-captured data.
In contrast, our proposed SMPL-X neural texture provides transformer blocks with coarse correspondences for cross-view matching, leading to better generalization. 
For in-the-wild images, we present the qualitative results of HumanRAM in the Supp. document.

\begin{figure}[ht]
    \centering
    \includegraphics[width=1\linewidth]{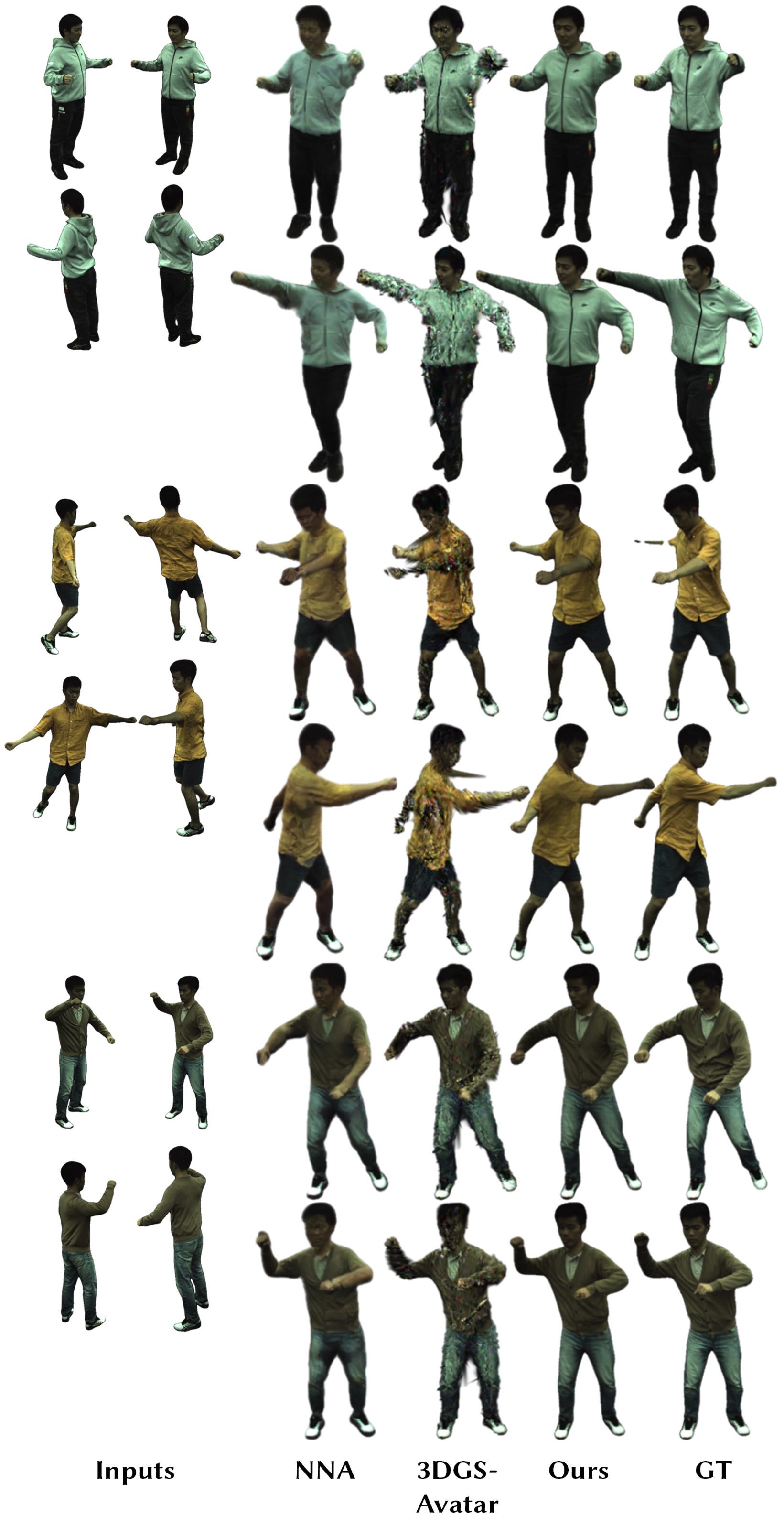}
    \caption{
    \textbf{Qualitative comparisons for multi-view animation on ZJUMoCap~\cite{peng2021neural}}. 
    We input 4 multi-view images of the unseen subject, and our method achieves a more photo-realistic rendering compared to other methods.}
    \label{fig:zjumocap_mv}
\end{figure}

\subsection{Comparison on Animation}
We compare HumanRAM with generalizable approaches including NNA~\cite{gao2023neural} and SHERF~\cite{hu2023sherf}, as well as a personalized approach 3DGS-Avatar~\cite{qian20233dgsavatar}, on ZJUMoCap~\cite{peng2021neural}.
We train HumanRAM and SHERF on THuman2.1~\cite{yu2021function4d} and ZJUMoCap~\cite{peng2021neural}. We use official weights for NNA since its training code has not been released.
For 3DGS-Avatar, we train it on input views and animate it with novel poses.
The evaluation is conducted on 100 randomly selected frames for each test subject.
The multi-view animation results are shown in Tab. \ref{tab:zjumocap_multi} and Fig. \ref{fig:zjumocap_mv}. The single-view results are shown in Tab. \ref{tab:zjumocap} and the Supp. document.
3DGS-Avatar requires a lengthy video to learn per-subject pose-dependent deformation.
However, when the data size is limited (single frame in our experiments), it is prone to overfitting the input images, leading to severe artifacts in novel views and poses.
NNA and SHERF learn a generalizable canonical avatar from the input image(s) and deform it to a novel pose using LBS wrapping.
Compared to 3DGS-Avatar, these generalizable methods achieve better animation results owing to the data-driven prior learning.
However, their canonical representation suffers from blurred textures and overfitting.
Besides, LBS wrapping tends to produce unnatural deformation in the underarm region.
Conversely, HumanRAM returns more realistic results in terms of quality and quantity thanks to the human structure prior learned through our dedicated designs.
We further present in-the-wild animation results in the Supp. document to demonstrate the generalization capacity of HumanRAM.
\begin{table}[!htbp]
\begin{center}
    \caption{
            \textbf{Quantitative comparison of reconstruction on ActorsHQ~\cite{isik2023humanrf}}. 
            All methods are evaluated directly on ActorsHQ without training or finetuning.
        }
    \label{tab:real_dataset}
    \begin{tabular}{ccccc}
        \toprule
         Metrics    & Ours & LVSM & GHG & LaRa \\ 
        \midrule
        PSNR$\uparrow$      &\textbf{25.47}     &20.25  &18.01   &19.98  \\
        SSIM$\uparrow$      &\textbf{0.9088}    &0.8023 &0.7922   &0.8177 \\
        LPIPS$\downarrow$   &\textbf{0.0350}    &0.0724 &0.0880   &0.0945 \\ 
        \bottomrule
        \end{tabular}
\end{center}
\end{table}
\begin{table}[!htbp]
\begin{center}
\caption{
            \textbf{Quantitative comparison of multi-view animation on ZJUMoCap~\cite{peng2021neural}}. 
            Metrics are computed on unseen subjects using the same crop manner as NNA~\cite{gao2023neural}.
        }
\label{tab:zjumocap_multi}
\begin{tabular}{cccc}
    \toprule
    Method &PSNR$\uparrow$ &SSIM$\uparrow$ &LPIPS$\downarrow$ \\
    \midrule
    \multicolumn{1}{l}{NNA} &21.29 &0.9369 &0.0530 \\
    \multicolumn{1}{l}{3DGS-Avatar} &18.50 &0.8367     &0.0499 \\
    \multicolumn{1}{l}{Ours} &\textbf{23.40}    &\textbf{0.9529} &\textbf{0.0252}\\
    \bottomrule
\end{tabular}
\end{center}
\end{table}

\subsection{Ablation Study}
\begin{figure}[!htbp]
    \centering
    \includegraphics[width=0.90\linewidth]{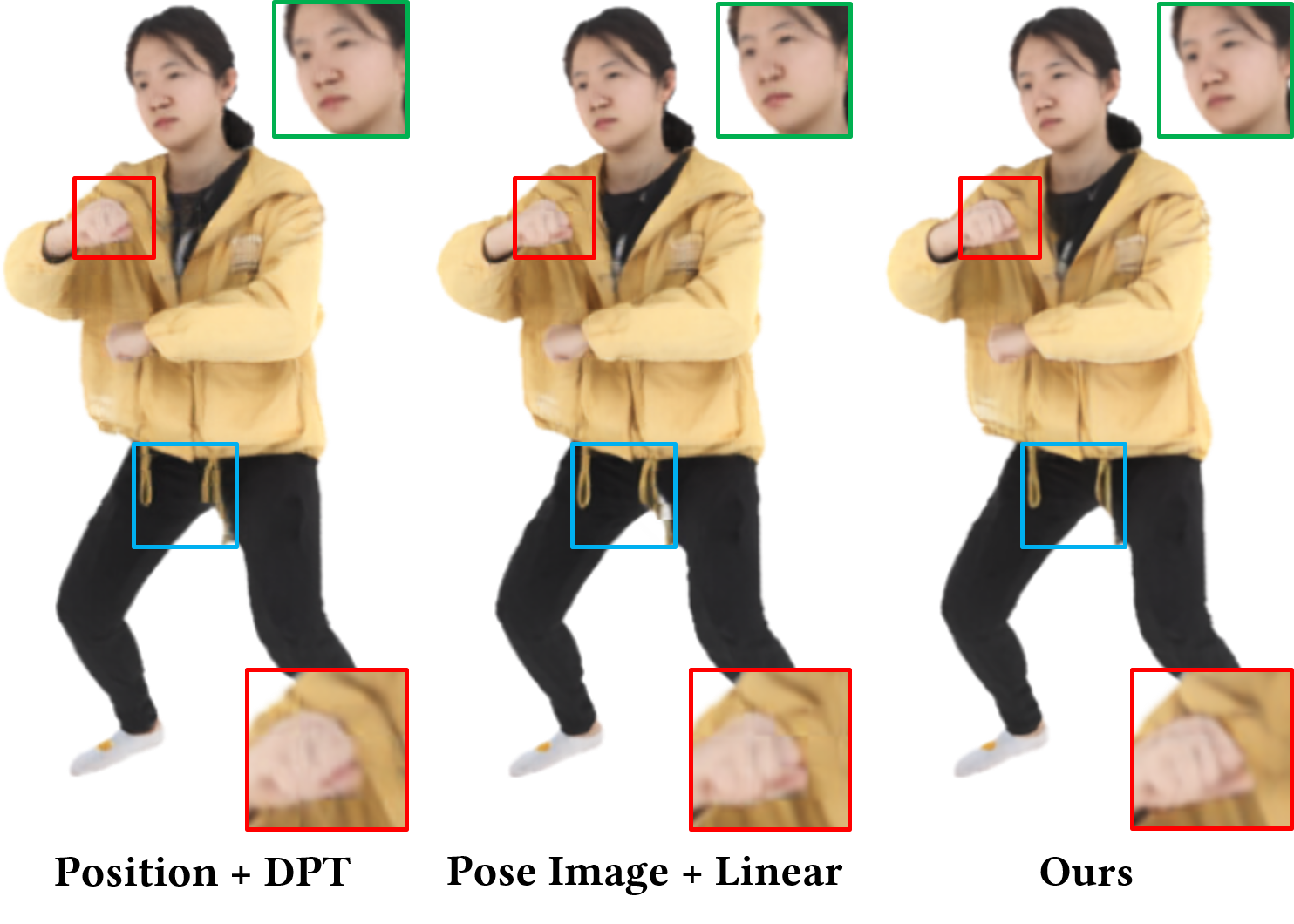}
    \caption{
    \textbf{Qualitative comparisons for ablations on proposed core components.} Compared with positions, pose image helps capture detailed structures (indicated by red and blue boxes). Furthermore, the DPT-based decoder helps reduce the patch-like artifacts in regions with severe self-occlusion (red box), thin structures (blue box), and face (green box). 
    }
    \label{fig:ablation}
\end{figure}

\paragraph{Core components.} 
We conduct ablation studies to evaluate the impact of our core components, i.e., \textit{Pose Image} and \textit{DPT-based Decoder}. The experiments are evaluated on the THuman2.1~\cite{yu2021function4d} dataset as shown in Tab. \ref{tab:ablation} and Fig. \ref{fig:ablation}. 
``Position'' means replacing the pose image with 3-dim position maps, and this ablation shows that such replacement decreases the performance, indicating the superiority of learnable neural texture. 
``Linear'' means replacing the DPT-based decoder with a linear layer used in vanilla LVSM.
Experiments show that the skip-connection and convolution operations in DPT are helpful in integrating information from multiple scales and neighboring patches, thus eliminating the patch-like artifacts and improving the overall visual quality.
\paragraph{Number of Views.} We evaluate the impact of view numbers on THuman2.1~\cite{yu2021function4d}. 
The model is directly evaluated on different numbers of input views without finetuning.
Tab. \ref{tab:ablation} shows the rendering quality increases with more input views, which aligns with the performance pattern reported in LVSM~\cite{jin2024lvsm}.
\begin{table}[ht]
\begin{center}
\caption{
            \textbf{Quantitative comparison of single-view animation on ZJUMoCap~\cite{peng2021neural}}. Metrics are computed on unseen poses and unseen subjects following SHERF~\cite{hu2023sherf}.
        }
\label{tab:zjumocap}
\begin{tabular}{ccccc}
    \toprule
    \multicolumn{1}{c}{} & \multicolumn{1}{l}{Method} & \multicolumn{1}{c}{PSNR$\uparrow$} & \multicolumn{1}{c}{SSIM$\uparrow$} & \multicolumn{1}{c}{LPIPS$\downarrow$} \\
    \midrule
    \multirow{3}{*}{\shortstack{Unseen\\Poses}} &\multicolumn{1}{l}{SHERF}  &18.56  &0.8760 &0.0501 \\
    &\multicolumn{1}{l}{3DGS-Avatar} &17.28 &0.8243 &0.0778 \\
    &\multicolumn{1}{l}{Ours} &\textbf{21.07}&\textbf{0.9152}&\textbf{0.0234} \\
    \midrule
    \multirow{3}{*}{\shortstack{Unseen\\Subjects}} & \multicolumn{1}{l}{SHERF} &17.80 &0.8768 &0.0536 \\
    & \multicolumn{1}{l}{3DGS-Avatar} &17.97 &0.8481 &0.0687 \\
    & \multicolumn{1}{l}{Ours} &\textbf{20.63}    &\textbf{0.9184} &\textbf{0.0250} \\
    \bottomrule
\end{tabular}
\end{center}
\end{table}
\begin{table}[ht]
\begin{center}
    \caption{
            \textbf{Ablation study on THuman2.1~\cite{yu2021function4d}}. We report PSNR, SSIM, and LPIPS to evaluate the contribution of proposed components and the impact of different input views. 
        }
    \label{tab:ablation}
    \begin{tabular}{lccc}
    \toprule
    \multicolumn{1}{c}{Method}           & \multicolumn{1}{c}{PSNR$\uparrow$} & \multicolumn{1}{c}{SSIM$\uparrow$} & \multicolumn{1}{c}{LPIPS$\downarrow$} \\
    \midrule
    Position + DPT  &29.32  &0.9443 &0.0197 \\
    Pose Image + Linear  &30.07  &0.9526 &0.0186 \\
    Ours (Pose Image + DPT, 4 views) &\textbf{30.34}  &\textbf{0.9535} &\textbf{0.0184} \\                   
    \midrule
    Ours (1 view)   &21.69 &0.8834 &0.0479 \\
    Ours (2 views)   &25.01 &0.9097 &0.0344 \\
    Ours (8 views)   &\textbf{32.34} &\textbf{0.9663} &\textbf{0.0150} \\
    \bottomrule
    \end{tabular}
\end{center}
\end{table}
\section{DISCUSSION}\label{discussion}
\paragraph{Conclusion.} We propose HumanRAM, a novel generalizable feed-forward model for human reconstruction and animation.
We integrate human reconstruction and animation into a unified framework by introducing pose conditions into large reconstruction models.
We introduce a shared SMPL-X neural texture and rasterize it onto input and target views to associate correspondences across different views and poses, enabling higher-quality reconstruction and realistic animation.
Overall, our method outperforms other state-of-the-art methods in terms of novel view and pose synthesis, both qualitatively and quantitatively.

\paragraph{Limitation.} 
Our method cannot handle high-resolution image inputs since the token number increases quadratically with the image resolution.
One possible solution is to transfer the inputs and outputs from the high-resolution RGB space to the compressed low-resolution latent space, like WonderLand~\cite{liang2024wonderland} and HumanSplat~\cite{pan2024humansplat}.
\begin{acks}
This work was partially supported by the following grants: National Key R\&D Program of China (No. 2024YFB2809105),
NSFC (No. U24B20154, No. 62172364),
Zhejiang Provincial Natural Science Foundation of China (No. LR25F020003),
Information Technology Center and State Key Lab of CAD\&CG, Zhejiang University,
and Research Grants Council of the Hong Kong Special Administrative Region, China (Project Reference Number: AoE/E-601/24-N).
\end{acks}
\bibliographystyle{ACM-Reference-Format}
\bibliography{ref}

\end{document}